\documentclass[twocolumn,superscriptaddress,preprintnumbers,amsmath,10pt,aps,prl,nobibnotes]{revtex4} 
\usepackage{amsmath}
\usepackage{amsfonts}
\usepackage{bm}
\usepackage{xcolor}
\usepackage{verbatim}
\usepackage{graphicx}
\usepackage{siunitx}
\usepackage{upgreek}
\usepackage{soul}

\begin{document}

\title{Photon-efficient optical tweezers via wavefront shaping}

\author{Un\.e~G.~B\=utait\.e}
\thanks{These authors contributed equally to this work}
\email{u.butaite@exeter.ac.uk}
\affiliation{School of Physics and Astronomy, University of Exeter, Exeter, EX4 4QL. UK.}

\author{Christina Sharp}
\thanks{These authors contributed equally to this work}
\email{u.butaite@exeter.ac.uk}
\affiliation{School of Physics and Astronomy, University of Exeter, Exeter, EX4 4QL. UK.}

\author{Michael Horodynski}
\affiliation{Institute for Theoretical Physics, Vienna University of Technology (TU Wien), A–1040 Vienna, Austria, EU.}

\author{Graham~M.~Gibson}
\affiliation{School of Physics and Astronomy, University of Glasgow, Glasgow, G12 8QQ, UK.}

\author{Miles~J.~Padgett}
\affiliation{School of Physics and Astronomy, University of Glasgow, Glasgow, G12 8QQ, UK.}

\author{Stefan Rotter}
\affiliation{Institute for Theoretical Physics, Vienna University of Technology (TU Wien), A–1040 Vienna, Austria, EU.}

\author{Jonathan~M.~Taylor}
\email{Jonathan.Taylor@glasgow.ac.uk}
\affiliation{School of Physics and Astronomy, University of Glasgow, Glasgow, G12 8QQ, UK.}

\author{David~B.~Phillips}
\email{d.phillips@exeter.ac.uk}
\affiliation{School of Physics and Astronomy, University of Exeter, Exeter, EX4 4QL. UK.}

\begin{abstract}
Optical tweezers enable non-contact trapping of micro-scale objects using light. Despite their widespread use, it is currently not known how tightly it is possible to three-dimensionally trap micro-particles with a given photon budget. Reaching this elusive limit would enable maximally-stiff particle trapping for precision measurements on the nanoscale, and photon-efficient tweezing of light-sensitive objects. Here we solve this problem by customising a trapping light field to suit a specific particle, with the aim of simultaneously optimising trap stiffness in all three dimensions. Initially taking a theoretical approach, we develop an efficient multi-parameter optimisation routine to design bespoke optical traps for a wide range of micro-particles. We show that the confinement volume of micro-spheres held in these sculpted traps can be reduced by one-to-two orders-of-magnitude in comparison to a conventional optical tweezer of the same power. We go on to conduct proof-of-principle experiments, and use a wavefront shaping inspired strategy to suppress the Brownian fluctuations of optically trapped micro-spheres in every direction concurrently, thus demonstrating order-of-magnitude reductions in their confinement volumes. Our work paves the way towards the fundamental limits of optical control over the mesoscopic realm.
\end{abstract}

\maketitle

\noindent The ability to remotely control the motion of small particles with laser light has become a key tool in a diverse range of scientific disciplines, from tests of fundamental physics to applications in the life sciences~\cite{marago2013optical,jones2015Optical,favre2019optical,bustamante2021optical}. The most widely used approach is the “optical tweezer”, first introduced by Arthur Ashkin in 1986~\cite{ashkin1986Observation}, enabling three-dimensional trapping of microscopic dielectric particles by tightly focusing a single Gaussian laser beam onto the target object. Since their conception, optical tweezers have found a multitude of applications. They have been used to reveal the bio-mechanics of molecular motors and protein-DNA interactions~\cite{block1990bead,heller2014optical}, to drive artificial micro-machines~\cite{grier2003revolution,butaite2019indirect}, to test the fundamental relationship between entropy and information~\cite{berut2012experimental,saha2021maximizing}, and to suspend nano-particles as their motion is cooled to the quantum ground state~\cite{delic2020cooling,
tebbenjohanns2021quantum}.

Given the widespread use of optical trapping, it may seem surprising that after more than 30 years, the most commonly employed spatial shape of laser beams used to create an optical tweezer is still the conventional Gaussian beam profile first suggested by Ashkin~\cite{ashkin1986Observation}. Indeed, Gaussian beams do come with many advantages: they are straight-forward to create and highly versatile -- operating in a broadly similar manner over a wide range of micro-particle sizes and shapes~\cite{jones2015Optical,simpson2014inhomogeneous}.

\begin{figure*}[t]
    \includegraphics{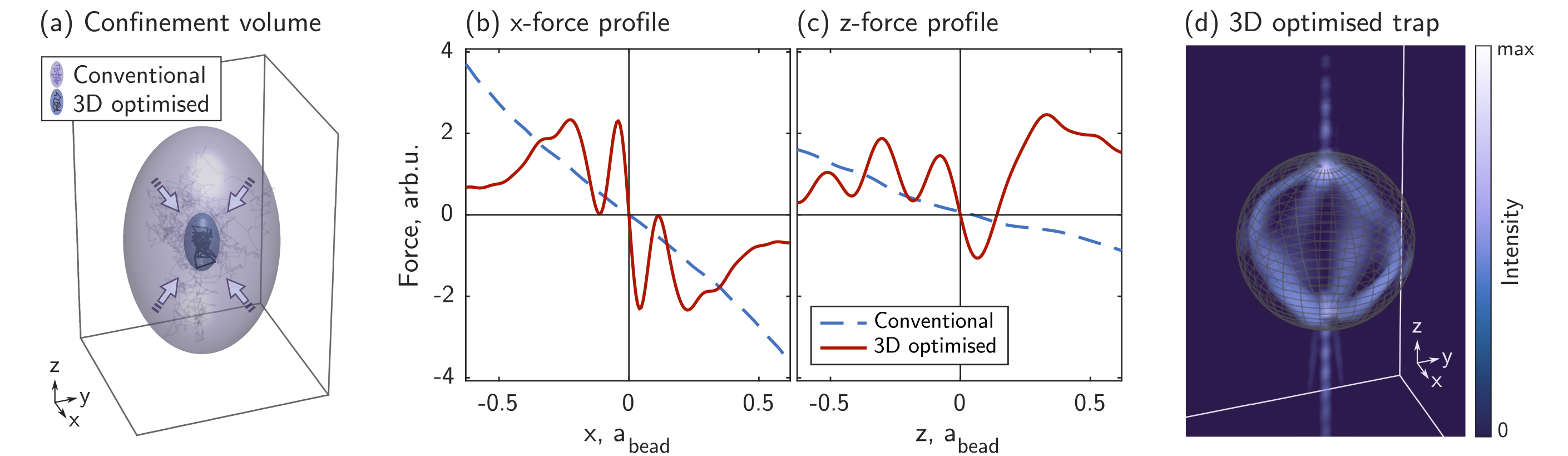}
    \caption{{\bf 3D optimised optical trapping.} (a) Confinement volume ($V_{\text{c}}$) of an optically trapped micro-particle in a conventional (violet) and an optimised (blue) optical trap, with simulated trajectories of the CoM shown. The confinement volume corresponding to the optimised trap is $\sim$74 times smaller. (b) Simulations showing the $x$-component of the optical force exerted on a micro-sphere as it is displaced laterally in the $x$-direction. (c) An equivalent plot to (b) but for the $z$-component of the optical force as the micro-sphere is displaced axially. A plot for the $y$-direction is available in SI\,\S10. (d) Schematic of the intensity of a 3D optimised optical trap (here we have accentuated the field inside the particle to present its features more clearly). In this figure we model a micro-sphere of \SI{5.56}{\micro\meter} in radius, with a refractive index of 1.48, immersed in water ($n = 1.3$), and illuminated from the negative $z$-direction with laser light of wavelength \SI{1.064}{\micro\meter}, through a numerical aperture (NA) of 1.25.} 
    \label{fig:force_curves}
\end{figure*}
However, this versatility comes at a cost: a Gaussian beam is typically not the optimal shape of light field to tightly trap a given micro-particle. This drawback is exacerbated for larger particles whose size is greater than the trapping wavelength. Such objects, which include many types of biological cells, are typically only weakly optically tweezed, suffering from a low trapping stiffness in a Gaussian beam. A straightforward way to overcome this issue is to crank up the laser power: doubling the intensity of the trapping light in turn doubles the trapping stiffness in all dimensions, thus confining a particle more tightly. Unfortunately, increasing the power focused onto the trapped object can lead to a variety of undesirable effects: excess photons can damage photosensitive biological systems~\cite{blazquez2019optical}, heat the particle and its local environment~\cite{peterman2003laser}, and increase decoherence effects in quantum ground state experiments~\cite{tebbenjohanns2021quantum}.

In this article we explore an alternate paradigm: we demonstrate to what extent it is possible to enhance 3D trap stiffness without increasing laser power, by instead tailoring the {\it spatial profile} of the laser beam. Optical fields have been previously shaped in a plethora of different ways to exert targeted optical forces and torques on micro-particles~\cite{friese1998optical,arlt2000generation,friese2001optically,ladavac2004sorting,roichman2008optical,dholakia2011shaping,padgett2011tweezers,brzobohaty2013experimental,woerdemann2013advanced,taylor2015enhanced,bezryadina2016optical,stilgoe2022controlled,hu2022structured}.
As far as enhancing trapping stiffness is concerned, various beam shapes have been tried in the past~\cite{friese1996determination,simpson1998optical,bowman2010particle,phillips2011optimizing,mcalinden2014accurate,singh2017particle}. An important step forward was taken by Taylor et al.\ in 2015, who demonstrated that carefully sculpting the incident optical field can deliver impressive {\it one-dimensional} lateral stiffness enhancements of trapped micro-spheres -- by up to a factor of $\sim$\,30 compared to conventional optical tweezers of the same power~\cite{taylor2015enhanced}. It is possible to identify globally optimum structured fields that accomplish these 1D stiffness enhancements using eigenvalue-based approaches~\cite{mazilu2011optical,taylor2017optimizing,agate2004femtosecond,ambichl2017focusing,horodynski2020optimal}. However, as we show in Supplementary Information (SI)\,\S1, these 1D trapping enhancements place no constraints on stiffness in other dimensions, and so do not guarantee that a particle is tightly trapped in 3D, or even stably trapped at all~\cite{volpe2023roadmap}. 

It is not straightforward to extend such eigenvalue-based strategies to enhance {\it multi-dimensional} optical trapping, since the stiffnesses along different directions are not independent. Consequently, the optimum 3D trapping field will not simply be a superposition of optimal trapping fields for the individual axes. Thus, despite multiple decades of research into optical trapping, an understanding of how to calculate the shape of `optimal' {\it three-dimensional} optical traps, or the level of multi-dimensional stiffness enhancement they may deliver, has remained out of reach~\cite{volpe2023roadmap}.

Here, we overcome these difficulties by designing bespoke trapping beams using an integrated multi-parameter optimisation strategy, guided efficiently to a solution by the information held within the Generalised Wigner Smith (GWS) operator~\cite{ambichl2017focusing,horodynski2020optimal}. Our approach allows all three dimensions to be considered simultaneously -- in terms of both stiffness enhancement and trap stability. Remarkably, we predict that custom-tailored trap shapes can hold micro-spheres up to a factor of 200 times more tightly than a Gaussian trap of equivalent power. To validate this concept experimentally, we develop a real-time optimisation routine that iteratively adapts the trapping wavefront to the shape of the particle. We experimentally demonstrate order-of-magnitude improvements in how tightly it is possible to hold micro-spheres in three dimensions. Our work establishes that dramatic gains in 3D optical trapping efficiency are possible by judiciously structuring light fields, and presents new theoretical and experimental routes to achieve them.\\

\noindent\textbf{Designing bespoke optical traps}\\
When submerged in a liquid, a trapped particle is constantly jostled around by collisions with surrounding molecules which are undergoing Brownian motion. But regardless of the direction in which the particle is displaced, the laser light of the trapping field is deflected to generate a near-Hookean optical restoring force, pulling the particle back towards its equilibrium position. For small displacements this force vector is given by ${\bm{f}^{\text{opt}}= -\boldsymbol{\kappa}\bm{\Delta r}}$, where $\boldsymbol{\kappa}$ is a 3$\times$3-element stiffness matrix encapsulating the translational trapping stiffness in any direction, and vector $\bm{\Delta r}$ describes the particle's 3D displacement from equilibrium~\cite{simpson2009thermal,phillips2012optically}.

Thermal motion of the fluid thus drives the particle's centre-of-mass (CoM) to stochastically explore a small region around its equilibrium position. To quantify the size of this region, we utilise the concept of the {\it confinement volume} $V_\text{c}$ -- see Fig.\,\ref{fig:force_curves}(a), also referred to as the thermal ellipsoid~\cite{phillips2012force}. The shape of the confinement volume is typically a prolate ellipsoid with its long axis parallel to the optical axis of the trapping beam -- reflecting the lower axial trapping stiffness arising from weaker intensity gradients in this direction. $V_\text{c}$ is given by:
\begin{equation}\label{Eq:thermEllipsoid}
    V_\text{c}=36\pi\sqrt{\frac{k_{\text{B}}^3T^3}{\kappa_x\kappa_y\kappa_z}},
\end{equation}
and is such that the probability of finding the CoM inside it is $p\sim 0.99$. Here $k_{\text{B}}$ is Boltzmann's constant, $T$ is the absolute temperature of the surrounding fluid, and $\kappa_{x,y,z}$ are the eigenvalues of $\boldsymbol{\kappa}$, which represent the stiffnesses of the optical trap along the principal axes of the thermal ellipsoid (see SI\,\S2). Our aim in this work is to find the shape of light fields that minimise the particle's confinement volume by maximising all eigenvalues of the stiffness matrix simultaneously.

We begin by investigating the level of 3D optical trapping enhancement possible from a theoretical perspective. We use the T-matrix formalism \cite{waterman1971symmetry,nieminen2007optical,barton1989theoretical} (see SI\,\S3) to model the interaction of a shaped incident light field $\bm u$ with a microscopic particle in an optical tweezers setup. The incident field coefficients held in the column vector $\bm u$ are defined at the pupil plane of the objective lens, expressed in the Bessel beam basis (see SI\,\S4), enabling us to explore the consequences of spatially shaping the intensity and wavefront of the trapping beam.

We seek the power-normalised complex values of $\bm u$ that correspond to a structured light field that maximises the trapping stiffness in three dimensions, while also stably holding the particle. To this end, we employ a constrained numerical optimisation routine, where we treat the desired relative stiffness in each dimension as a tunable parameter -- yielding flexibility in the final aspect ratio of the particle's confinement volume (see Methods and SI\,\S5). A large number of optimisation variables $N$ (i.e.\ elements of $\bm u$) is required in order to shape the light effectively. This makes the problem computationally demanding, with previous examples of numerical 1-D trapping field optimisation taking on the order of days to converge~\cite{taylor2015enhanced,butaite2020enhanced}. Therefore we hone the efficiency of our optimisation routine to significantly speed up this process. 

\begin{figure*}[t] \includegraphics{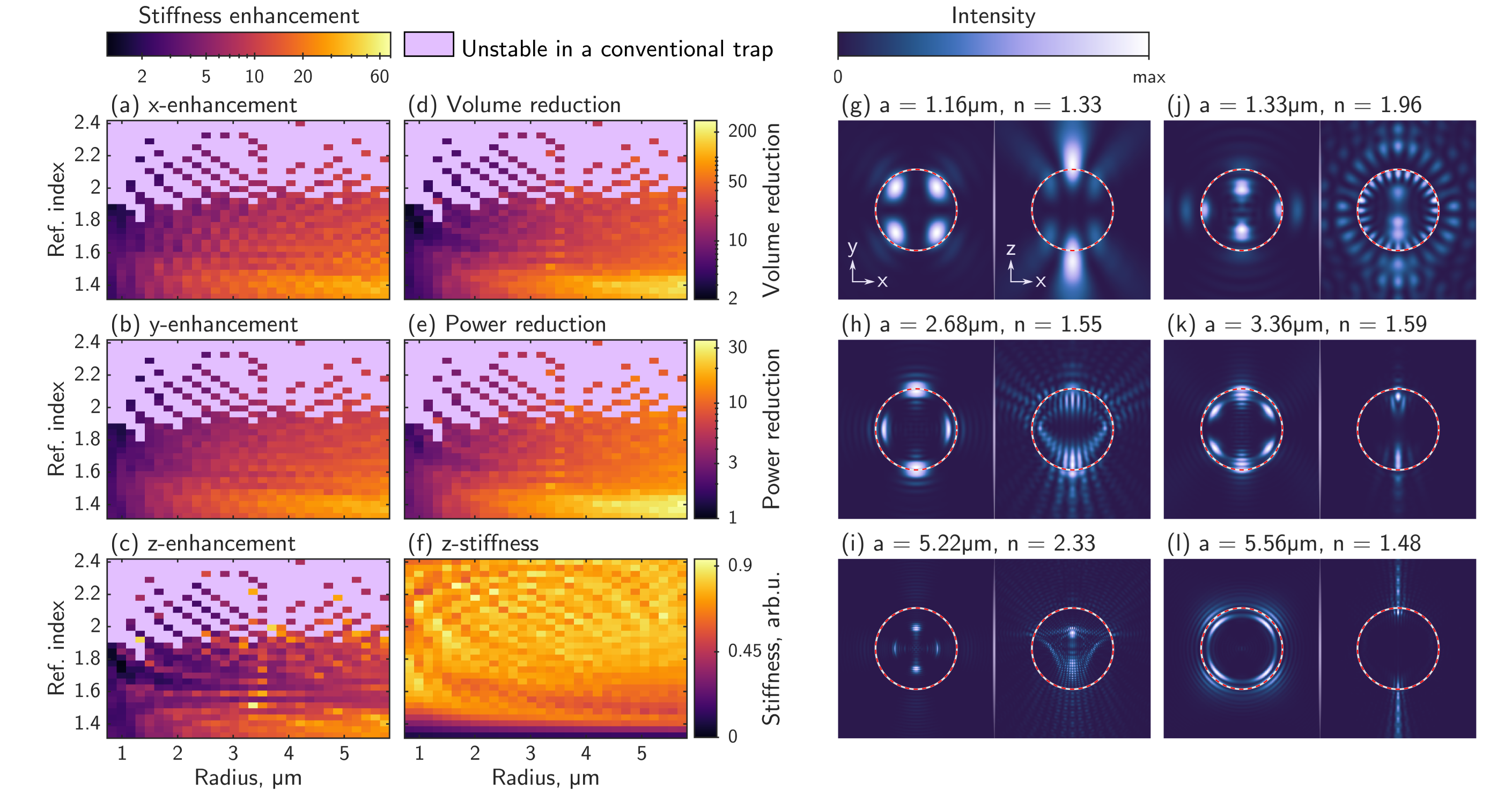}
    \caption{{\bf Exploring 3D optical trap enhancement as a function of micro-sphere size and refractive index.} (a-c) Simulated trapping stiffness enhancement (when compared to a conventional optical trap) achieved simultaneously in each dimension, as a function of micro-sphere radius and refractive index. (d) Volume and (e) power reduction resulting from these stiffness enhancements. Lavender shaded areas indicate particles which cannot be stably trapped along the axial direction in conventional optical tweezers. Note that (a-e) are plotted on a logarithmic scale to better reveal detail. (f) Simulated $z$-stiffness achieved with our optimised traps; note that by design $x,y$-stiffness is 3.2 times greater than the $z$-stiffness (see SI\,\S5) and follows the exact same trend. (g-l) Examples of optimised trap intensity cross-sections; the light is propagating in the positive $z$-direction, and the white-red outline indicates the edges of the micro-sphere. Here we used the same parameters as in Fig.\,\ref{fig:force_curves}.}
    \label{fig:heatmapFig}
\end{figure*}

The cornerstone of our approach is the GWS operators for optical force ($\bm{Q}$) and trapping stiffness ($\bm{K}$)~\cite{wigner1955lower,smith1960lifetime,ambichl2017focusing,horodynski2020optimal}. These operators reduce the evaluation of force and stiffness along a particular direction to a single matrix equation. For example, along the $x$-dimension we have:
\begin{align}
    f_x^\mathrm{opt} &= \bm{u}^\dagger \bm{Q}_x \bm{u} = \bm{u}^\dagger \left(-\text{i}\bm{S}^\dagger \partial_x \bm{S}\right) \bm{u}, \label{eq:force}\\
    \kappa_x &= \bm{u}^\dagger \bm{K}_x \bm{u} = \bm{u}^\dagger \left(-\partial_x \bm{Q}_x\right) \bm{u}, \label{eq:stiffness}
\end{align}
where $\bm{S}$ is the scattering matrix describing how the light interacts with the particle (i.e.\ which incident light modes are scattered into which outgoing modes), $\partial_x$ indicates a partial derivative with respect to the $x$-position of the particle, and $^\dagger$ indicates a conjugate transpose. Equivalent expressions can be written down for any other degree of freedom, including rotations. Crucially, $\bm{Q}$ and $\bm{K}$ only encapsulate properties of the particle, and are independent of the incident field. As such, they only need to be evaluated once, before the optimisation routine commences, significantly cutting down on computational time. Expressions for $f^\mathrm{opt}$ and $\kappa$ are also readily differentiable with respect to $\bm{u}$, providing access to analytical expressions for gradients and Hessians, thus further speeding up the optimisation. See SI\,\S1,6-8 for more detail on $\bm{S}$, $\bm{Q}$, and $\bm{K}$, including our derivations of analytical expressions for calculating the derivatives in Eqs.\,\ref{eq:force}--\ref{eq:stiffness}. This approach reduces the time-scale for a single trap design from days to minutes (see SI\,\S9), allowing us to explore the enhancements achievable over a wide range of different particles.\\

\noindent\textbf{Theoretical trapping enhancements}\\
An example of one such optimised trap can be seen in Fig.~\ref{fig:force_curves} -- here designed for a micro-sphere of diameter $\sim$10 times larger than the laser wavelength. Our modelling shows that the achieved confinement volume is $\sim$74 times smaller than that of a conventional optical trap carrying the same power -- as seen in Fig.~\ref{fig:force_curves}(a). Figure~\ref{fig:force_curves}(b-c) presents optical force curves of the optimised trap, indicating that the trapping stiffness has been increased by factors of $\sim$19 and $\sim$15 in the $x$ and $z$-directions respectively. In Fig.~\ref{fig:force_curves}(d) we see that inside the particle the optimised optical field tracks around the micro-sphere's surface -- in stark contrast to a conventional Gaussian beam which typically focuses at its centre. This optimised field shape can be understood by considering that momentum transfer between laser light and a trapped particle can only take place where there is a spatial gradient in refractive index -- i.e.\ at the interface between the particle and the surrounding medium~\cite{phillips2014shape} (for a non-absorbing, homogeneous and isotropic particle, as in this case). Thus our optimiser achieves three dimensional trapping stiffness enhancement by boosting the intensity of light at the particle's boundaries, as well as simultaneously ensuring a stable trap by balancing momentum transfer due to specular reflections from the particle.
Figure~\ref{fig:heatmapFig} shows the theoretical optical trapping enhancements possible for 900 scenarios involving spherical particles of different sizes (in the range 0.8\,--\,\SI{5.7}{\micro\meter} in radius) and refractive indices ($n =$ 1.33\,--\,2.4). We optimise the complex amplitude of $N = 820$ spatial modes within each trapping beam. Figures~\ref{fig:heatmapFig}(g-l) show examples of the intensities of these bespoke optical traps in transverse and axial cross-sections through the particles. We compare the theoretical performance of these optimised optical traps against a conventional optical trap of the same NA in two ways: we show the directional stiffness enhancement as the factor of improvement in the eigenvalues of $\boldsymbol\kappa$ in Fig.~\ref{fig:heatmapFig}(a-c), and the confinement volume reduction $V^{\text{rel}}_{\text{c}}$ as the factor of improvement in $V_\text{c}$ in Fig.~\ref{fig:heatmapFig}(d).

Our simulations indicate that it is possible to achieve very substantial 3D trapping enhancements -- in some cases of up to a 200-fold reduction in the confinement volume, and with stiffness enhancements exceeding $\sim$20 in all dimensions simultaneously for a broad range of micro-sphere sizes and refractive indices. Equivalently, these reductions in particle confinement volume can be translated into improvements in relative trapping efficiency $\eta^{\text{rel}}$: the factor of reduction in trapping power needed to hold a particle as tightly as a conventional optical trap, where $\eta^{\text{rel}}= \left(V^{\text{rel}}_{\text{c}}\right)^{2/3}$ (see SI\,\S11). In Fig.~\ref{fig:heatmapFig}(e) we see that $\eta^{\text{rel}}$ is over a factor of ten for a wide range of particles -- predicting that the same trapping performance as a conventional optical tweezer can be achieved using less than one-tenth of the laser power. We find that these enhancements are highly repeatable, irrespective of the choice of initial values used for the complex optimisation variables $\bm u$ (see SI\,\S12). This suggests we may be finding solutions that are close to the global optimum, given the set of constraints that we have enforced.

\begin{figure*}[t]
    \includegraphics{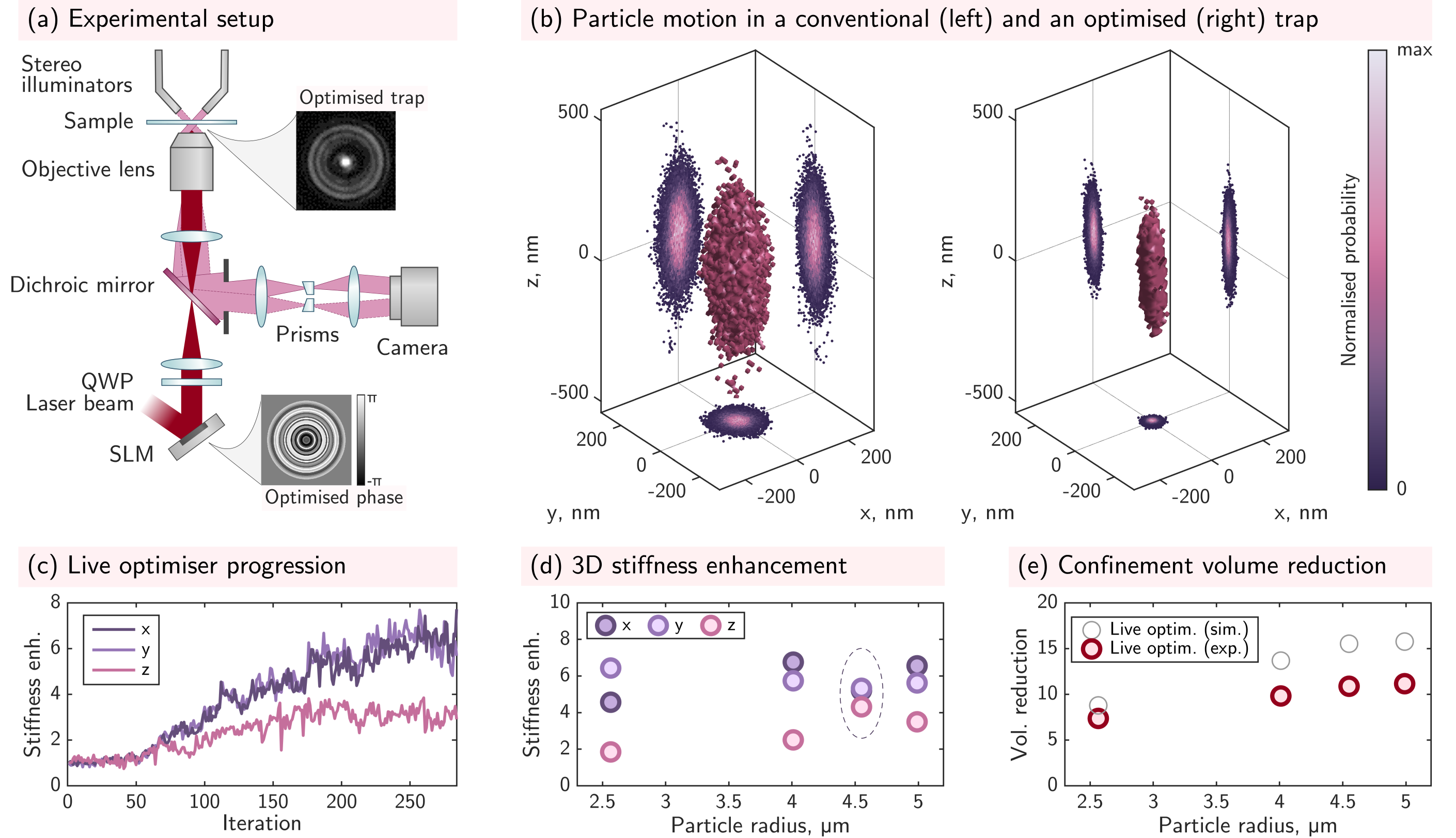}
    \caption{{\bf Experimental results of live-optimised optical trapping.} (a) Schematic of the holographic optical tweezers system used for the experiments presented here. (b) Experimental data showing the thermal volume explored by a \SI{5}{\micro\meter} radius silica micro-sphere, with projections of 2D position probability distributions of the CoM. (c) Example of the progression of the live optimiser in experiment. (d) Typical stiffness enhancements obtained for a range of particle radii (as quoted by manufacturers) and (e) corresponding volume reduction factors. We also simulate the ideal performance of live-optimisation, points shown in grey in (e). See SI\,\S21 for a discussion of errors.}
    \label{fig:expFig}
\end{figure*}

A notable feature of our method is that it predicts an extension of the range of particle parameters over which optical trapping is possible: conventional optical tweezers fail to create stable $z$-equilibria for high refractive index particles, while we see in Fig.~\ref{fig:heatmapFig}(f) that stiff and stable optical trapping becomes possible in this high-index regime by shaping the structure of the trapping field. Another dominant trend in these results is that large particles with low refractive indices benefit the most. The enhancements are lower for small dielectric particles ($<$\SI{1}{\micro\meter} radii) because the diffraction limit restricts the extent to which the trapping light can be shaped within the footprint of the particle.

Driven by typical experimental capabilities, we also investigate enhanced optical trapping using phase-only beam shaping, rather than the intensity and phase control employed so far. We reformulate our numerical optimisation routine to incorporate this constraint -- see SI\,\S13. We observe that the thermal volume compression achieved by phase-only optimised traps is typically $\sim$\,60\,\% of that achieved with full-field light shaping, across the 900 scenarios considered. Nonetheless, our analysis shows substantial trapping enhancements over conventional optical tweezers are still possible with phase-only light modulation.\\

\noindent\textbf{Live-optimised 3D optical trapping}\\
Encouraged by the predictions of our fully vectorial 3D model, we now investigate the experimental implementation of optimised 3D optical trapping. We employ holographic optical tweezers and a 3D particle tracking platform, shown schematically in Fig.~\ref{fig:expFig}(a). The trapping field is shaped with a liquid crystal spatial light modulator (SLM), conjugated with the pupil of the objective lens, and the 3D motion of trapped particles, immersed in water, is tracked in real-time (at up to \SI{1}{\kilo\hertz}) using high-speed stereo-microscopy, delivering nano-metric precision in three dimensions~\cite{bowman2010particle,hay2014four}. See Methods for more detail.

A number of specific challenges arise when seeking to apply a numerically pre-designed trapping field in a real-world experiment.
Commercially available colloids exhibit a statistical variation in size, and our modelling shows (SI\,\S14) that the optimal trap shape is highly dependent on the exact physical properties of the target particle. The volumetric trapping fields must also be generated with extremely high fidelity, since the mechanical response of the particle is highly sensitive to small variations in the applied trapping field. The creation of pre-designed fields is influenced by factors including: the imperfect response of the SLM~\cite{moser2019model,kupianskyi2023high}, the precise action of the objective lens~\cite{gu1996effect,sheppard1988aberrations,he2021vectorial}, and the configuration of the oil-glass-water interfaces that light passes through on its way to the target particle~\cite{iwaniuk2011correcting} -- even after \textit{in situ} aberration correction is performed with the SLM~\cite{vcivzmar2010situ}. The combined effect of the above factors means that employing pre-designed optimised traps is very challenging.

We therefore devise a new experimental strategy compatible with these challenges: we take inspiration from the field of {\it wavefront shaping} which has emerged as a powerful way to optimise coherent light transport in unknown complex scattering environments~\cite{vellekoop2007focusing,gigan2022roadmap,cao2022shaping}.
We implement a {\it live} iterative optimisation routine that automatically tailors the trapping field to the (unknown) properties of a target particle. The optimisation is conducted using real-time measurements of the three-dimensional trapping strength, as inferred from the stochastic trajectory of the particle. Crucially, our approach does not require \textit{a priori} knowledge of the properties of the target particle, or the optical characteristics of optical tweezers platform itself, since the optimisation is conducted based purely on experimentally-measured metrics. See Methods for a detailed description of our algorithm. 

Figure~\ref{fig:expFig}(b) displays measured point clouds representing the confinement volume explored by the CoM of a \SI{5}{\micro\meter} radius silica micro-sphere trapped in water using conventional optical tweezers (left) and an optimised trap (right). We see that the thermal ellipsoid has shrunk in all dimensions -- in this case reducing in volume by a factor of $V^{\text{rel}}_{\text{c}}\sim13$. We validate our live optimisation strategy on silica micro-spheres ranging in size from 2.5-\SI{5}{\micro\meter} in radius. Figure~\ref{fig:expFig}(c) shows a typical progression of the live optimiser.

Figure~\ref{fig:expFig}(d) shows examples of trapping stiffness enhancements for these different micro-sphere sizes. In all cases we observe that stiffness enhancement factors in the $x$ and $y$ dimensions ($\sim6-8$) exceed those in the $z$ dimension ($\sim2-3$). Cases where the $z$ stiffness enhancement is larger result in a concomitant reduction in the $x$ and $y$ enhancements, as highlighted for the trap optimisation targeting the \SI{4.5}{\micro\meter} radius micro-sphere (points enclosed by a dashed ellipse).

Figure~\ref{fig:expFig}(e) shows the 3D confinement volume reduction factors ($V^{\text{rel}}_{\text{c}}$) for each micro-sphere size. Our experimentally optimised traps deliver about an order of magnitude reduction in the confinement volume across all tested particles. This corresponds to an improvement in trapping efficiency by a factor of $\sim5$: i.e.\ the same 3D trapping stiffness as conventional optical tweezers can be achieved using only $\sim20\%$ of the laser intensity. Traps optimised for larger micro-spheres tend to yield higher enhancements, consistent with our expectations from theory and simulations. For the smallest and largest micro-spheres, we perform five optimisation runs on different particles within the sample to demonstrate the repeatability of our live-optimisation strategy (see SI\,\S15), despite variation in particle size, and the presence of significant measurement noise.\\

\noindent\textbf{Discussion}\\
We have shown the first experimental demonstrations of optical traps specifically customised to enhance micro-particle confinement in all three dimensions simultaneously. For the experimentally tested range of silica micro-spheres, our modelling of ideal pre-designed traps predicts that it is possible to use phase-only optimisation to suppress confinement volumes by factors of ${V^{\text{rel}}_{\text{c}}\approx15-30}$ (see SI\,\S16.). Our proof-of-principle experiments reach $\sim33-50\%$ of these values (${V^{\text{rel}}_{\text{c}}\approx8-13}$). We attribute these differences mainly to the reduced search space available to the live optimiser (see Methods for a detailed discussion of this effect).

In order to better understand the limitations of our live optimisation strategy, we simulate its performance when stiffness measurements are subjected to the level of noise found in our experiments (for experimental noise analysis see SI\,\S17.). Figure~\ref{fig:expFig}(e) shows the anticipated reduction in confinement volume based on these simulations (grey circles). By comparison, our experiments reach $\sim$\,70-90\,\% of these values. Residual differences are due to real-world SLM diffraction losses that occur as the complexity of the displayed patterns increases (see SI\,\S20), and the presence of minor system aberrations that break the circular symmetry of the trapping field, neither of which feature in our simulations.

In the future it may be possible to develop faster and more sophisticated live-optimisation algorithms with improved resilience to noise~\cite{mastiani2021noise,mididoddi2023threading}. Furthermore, we anticipate it will become feasible to directly experimentally deploy pre-optimised traps following high-fidelity system and sample calibration -- so that the trap design algorithm accurately captures the capabilities of the experimental platform. SI\,\S16 shows a direct comparison of our experimental results with all simulated optimisation approaches, highlighting the future potential of such customised 3D optical trapping.

We now consider the advantages and trade-offs of enhanced optical trapping more broadly. Our optimised trap design algorithm is efficient and versatile: capable of full-field or phase-only optimisation, compatible with microscopes of any numerical aperture up to a solid angle of $4\pi$ (thus including counter-propagating dual-beam trapping~\cite{ashkin1970acceleration,thalhammer2011optical}), and can be readily extended to optimise beams with spatially varying polarisation, and to multi-spectral light control. Our strategy also allows the aspect-ratio of the confinement volume to be freely tuned~\cite{bowman2010particle} (as shown in SI\,\S5.). Additional optimisation criteria, such as prescribed optical forces, or optical torques about any axis, can also be specified using the GWS operators~\cite{horodynski2020optimal,stilgoe2022controlled,hu2022structured}. 

Here we have focused on the discovery of new traps for homogeneous and isotropic micro-spheres. However, these design techniques can be applied to optimise traps for particles of arbitrary geometry by first pre-calculating the shape's T-matrix -- in these cases, the optimised trapping fields will depend upon particle orientation. We have shown how high refractive index particles, previously considered `untrappable', have the potential to be stably held using appropriately shaped beams. Furthermore, our design algorithm can also generate new forms of optimised `bottle-beam' or `dark' optical tweezers~\cite{arlt2000generation,xiao2021efficient}, capable of stably holding objects of lower refractive index than the surrounding medium. Current dark trap design methods struggle with extended low-index particles that are significantly larger than the diffraction limit~\cite{melo2020optical} -- a challenge our optimiser can overcome, as shown in SI\,\S18. 

It is important to note that trapping stiffness enhancements tend to come at the expense of a reduction in the energy barrier preventing a particle from escaping the trap -- as also observed previously for 1D optimised traps~\cite{taylor2015enhanced}. This effect is noticeable in Fig.~\ref{fig:force_curves}(b-c), where the stiffness at the origin is many times higher for the optimised trap (red curve), yet the trapping range, as well as the maximum restoring force, are reduced (see also extended plots in SI\,\S10.) In practice, this means that the beam must carry enough power to counteract the thermal motion of the particle and prevent it from `jumping out' of the trap, and this effect becomes more acute for traps optimised for particles of high refractive index (see SI\,\S19). 

Our concept relies on being able to shape the incident field across the footprint of the target object. Therefore, dielectric particles of a diameter close to the diffraction limit and smaller do not benefit from dramatic 3D stiffness enhancements over conventional optical tweezers -- although we note that sub-diffraction limited super-oscillating beams may offer new opportunities for tightly trapping smaller particles~\cite{singh2017particle}. In addition, lossy metallic particles cannot presently be treated by our fast optimiser in its current form, due to its reliance on flux-conservation. Recently, efficient single-parameter numerical optimisation of force or torque on metallic nano-particles was demonstrated~\cite{lee2017computational}, pointing to a way forward in this regime.

Finally, we note that holographic beam shaping has previously been used to adapt the 3D intensity of light to match the shape of trapped particles~\cite{kim2017tomographic} -- an approach capable of holding irregularly shaped objects at desired positions and orientations, which can be viewed as optimising the trapping {\it stability}. Our concept is fundamentally different -- we aim to optimise trapping {\it stiffness}, which entails shaping light fields to create high intensity gradients on the boundaries of a homogeneous particle, rather than projecting uniform intensity throughout its volume. Recent work has also begun to explore new ways to identify regions of high refractive index gradients within large inhomogeneous particles, in order to exert higher optical forces by focusing light onto these areas~\cite{landenberger2021nonblind}.\\

\noindent\textbf{Conclusions}\\
In summary, we have shown there is plenty of scope to enhance optical trapping through wavefront shaping, and that order of magnitude improvements in 3D particle confinement are now within reach. In the future we envision a union of customised light fields applied in concert with specifically engineered micro-particles will lead to new ultra-stiff and high-force optical traps with specialist capabilities~\cite{friese2001optically,hu2008antireflection,jannasch2012nanonewton,phillips2014shape}. The areas of optical trapping that we expect to benefit most from these advances are those that require ultra-precise manipulation of micro-particles, or feature samples intolerant of high optical intensities. Examples include optical traps used to isolate particles as their motion is cooled to the quantum ground state -- where excess photons result in additional heating and quantum decoherence~\cite{piotrowski2023simultaneous}, precision positioning of microscopic sensors~\cite{millen2014nanoscale}, automated optical assembly of micro-scale structures~\cite{melzer2021assembly}, and the study of photo-sensitive biological systems~\cite{blazquez2019optical}. The work we have presented here offers new routes towards the optimal transfer of momentum from the photonic to the micro-mechanical regime -- pushing towards the fundamental limits of passive far-field optical control over matter.

\section{Methods}
\noindent\textbf{Simulated optimisation of optical traps}\\
Our modelling framework is built upon the freely available Optical Tweezers Toolbox (OTT)~\cite{nieminen2007optical}, with several custom modifications to integrate the GWS operators which allow fast optical force and stiffness calculations. For the optimisation itself we used MATLAB's `fmincon' function with the `interior-point' algorithm, which is designed for non-linear constrained optimisation. We note that the `fmincon' function is not capable of dealing with complex numbers, so we split $\bm{u}$ into its real and imaginary parts to perform the optimisation. 

We set $\kappa_x$ as the objective function to be maximised, and then, to make sure that the solution light field possess the desired stiffness and stability requirements we set the following constraints. Mimicking the properties of the conventional optical trap we want the transverse stiffness to be isotropic, so we require that $\kappa_y = \kappa_x$. And to further cement this property and avoid solutions which preferentially treat the $y=x$ or $y=-x$ direction, we also require that $\kappa_{y=x,y=-x} = \kappa_x$. For the $z$-direction we follow the symmetries of a diffraction limited spot (i.e.\ lower stiffness axial trapping) and require that $\kappa_z=\kappa_x/3.2$ . We do note, however, that any desired aspect ratio between the stiffnesses along different dimensions can be specified (see SI\,\S5). We further require that there is no optical force acting on the particle at the origin -- $f_{x,y,z}(\bm{0})=0$ -- so as to ensure existence of a stable equilibrium. And lastly, we require normalised power such that $|\bm{u}|^2 = 1$. We also note that some of these constraints can be removed if the basis in which $\bm{u}$ is expressed is itself limited to certain symmetries as, for example, is the case for our phase-only optimiser.\\

\noindent\textbf{Live experimental optimisation}\\
In our experimental optimisation routine, we aim to minimise the number of optimisation variables $N$ in order to converge to a solution as rapidly as possible. We achieve this by exploiting knowledge of the symmetries of target particles and the optical system itself. We mimic the Bessel basis used in simulations by splitting our SLM screen into $N$ evenly radially-spaced rings, and aim to determine the relative phase that should be imparted to light reflecting from each ring. In this geometry, use of circularly-polarised light limits the search space to cylindrically symmetric fields -- well matched to the spherically shaped target particles. 

The phase of each ring is optimised as follows. At the start of each iteration, the optimiser randomly selects half of the $N$ rings, and adds a small phase change $\Delta\phi$ to them. The particle's CoM is then tracked for a time $\Delta t$ to accumulate enough data for evaluating the trap stiffness $\bm{\kappa}_{+\Delta\phi}$ (using the Equipartition theorem~\cite{jones2015Optical}). Next, $\Delta\phi$ is subtracted from the same set of rings, and $\bm{\kappa}_{-\Delta\phi}$ is evaluated. We then also perform the stiffness evaluation on the initial phase configuration -- this way we avoid the optimiser getting stuck in a noise-induced false `high-stiffness' configuration. From the three phase configurations the one with the best stiffness is selected -- this completes one iteration, and the process is then repeated. The `best field' within each iteration is determined as the one that increases $\kappa_{x,y}$ and does not decrease $\kappa_z$ (or vice versa), and we rely on the symmetries of the rings, the spherical particle itself, and circularly polarised light to ensure that $\kappa_y$ does not diverge significantly from $\kappa_x$. 

Substantial measurement times $\Delta t$ are required for precise and accurate measurements of trap stiffness, due to the stochastic nature of Brownian motion~\cite{perez2018high}. In order not to wait too long per iteration, we use the shortest time that still enables successful optimisation, which means the convergence process is inherently noisy as shown in Fig.~\ref{fig:expFig}(c). The phase step size $\Delta\phi$ tested at each iteration must be small enough to ensure that any trial fields projected onto the particle will not eject it from the trap and end the experiment. At the same time, $\Delta\phi$ needs to be large enough so that the change in stiffness is detectable above the thermal noise, for the chosen measurement time $\Delta t$. Typical values in our experiments were $\Delta\phi=\pi/10$, $N=30$, and $\Delta t=\SI{10}{\second}$. We also avoid abrupt jumps in the phase profile displayed on the SLM (which lead to higher loss in diffraction efficiency) by linearly interpolating the phase between the $N$ rings.

Our live-optimisation algorithm is designed to be relatively tolerant of noisy experimental measurements~\cite{mididoddi2023threading}. However, in comparison to the optimiser used to pre-design optical traps, live optimisation has a reduced search space, due to the need to maintain trap stability at all steps throughout the optimisation pathway. This constraint limits the enhancements that live-optimisation can deliver when compared to the pre-designed traps, even in noiseless conditions (see SI \S16).\\

\noindent\textbf{Holographic optical tweezers with 3D tracking}\\
Our holographic optical tweezers setup is schematically detailed in Fig.\,\ref{fig:expFig}(a). It is based on a modified version of the ``cube" optical tweezers platform presented in ref.~\cite{gibson2012compact}. A \SI{1064}{\nano\meter} continuous-wave diode-pumped solid-state laser (Laser Quantum: VentusIR, \SI{3}{\watt}) is expanded to fill a liquid crystal spatial light modulator (Boulder Nonlinear Systems: XY-series, 512 × 512 resolution), which is in turn re-imaged onto the back of a \SI{1.3}{NA} \SI{100}{x} oil-immersion objective (Olympus) using a 4$f$-imaging system. A sample slide holding a dilute suspension of silica micro-spheres (microParticles Gmbh) is placed in the front focal plane of the objective, where the micro-spheres can be manipulated using wavefront-shaped optical traps. 

We implement stereoscopic vision for 3D particle tracking. The sample is back-illuminated with two red LED sources, forming twin views of the sample from different angles. The two images are collected by the same objective lens and later passed through two spatially adjacent prisms, positioned in the Fourier plane of the sample, to separate the two `eyes' of the stereo-vision system. Finally the two spatially separated views of the sample are imaged side-by-side onto a high-speed camera (Mikrotron, EoSens CL). 2D centre-of-symmetry based real-time tracking in each image enables reconstruction of 3D micro-sphere trajectories using parallax~\cite{bowman2010particle}, with nanometric axial precision~\cite{hay2014four} (see SI\,\S21). The system is operated using the LabVIEW based ``Red Tweezers" software~\cite{bowman2014red}, which is modified to incorporate stereo-vision 3D tracking. For this work we made further changes to implement our custom live-optimisation routine.

The optical traps used in the experiments presented here were generated using the first diffraction order from the light shaped by the SLM. We found that using the full NA of the objective lens resulted in higher levels of aberration. To combat this, we reduce the NA of our system from 1.3 to 1.13 by including a circular aperture on the phase masks displayed on the SLM -- thus only light reflecting from within the aperture is transmitted to the first diffraction order, cutting out light from the edges of the SLM. This circular aperture was used for all Gaussian and optimised trapping experiments to ensure an identical NA in all cases.

\section{Acknowledgements}
DBP acknowledges the financial support from the European Research Council (Grant no.\ 804626), and the Royal Academy of Engineering. UGB acknowledges financial support from EPSRC (EP/N509668/1). MH and SR are supported by the Austrian Science Fund (FWF) through project No. P32300, and acknowledge access to the Vienna Scientific Cluster (VSC). JMT acknowledges the Royal Society of Edinburgh (sabbatical grant). We used the Scientific Colour Map suite \cite{crameri2020misuse,crameri2021scientific} for our figures; Fig.\,\ref{fig:force_curves}(d) was produced using the Matlab VOXview function \cite{tim2021voxview}.

\section{Contributions}
UGB, JMT and DBP conceived the idea for the project, and developed it with all other authors. DBP and JMT supervised the project. UGB led the development of the optimiser and performed all simulations, with support from CS. MH and SR developed the theory of the GWS operators and derived the gradients and Hessians used in the optimisation. JMT derived the analytical expressions for the differential of the scattering matrix. GMG and CS built the optical setup, with support from DBP and MJP. UGB and CS modified the experimental control software. CS performed the experiments and data analysis with support from UGB and DBP. UGB, CS, JMT and DBP wrote the manuscript with editorial input from all other authors.

\bibliographystyle{naturemag}
\bibliography{MasterBib.bib}

\begin{thebibliography}{10}
\expandafter\ifx\csname url\endcsname\relax
  \def\url#1{\texttt{#1}}\fi
\expandafter\ifx\csname urlprefix\endcsname\relax\def\urlprefix{URL }\fi
\providecommand{\bibinfo}[2]{#2}
\providecommand{\eprint}[2][]{\url{#2}}

\bibitem{marago2013optical}
\bibinfo{author}{Marag{\`o}, O.~M.}, \bibinfo{author}{Jones, P.~H.},
  \bibinfo{author}{Gucciardi, P.~G.}, \bibinfo{author}{Volpe, G.} \&
  \bibinfo{author}{Ferrari, A.~C.}
\newblock \bibinfo{title}{Optical trapping and manipulation of nanostructures}.
\newblock \emph{\bibinfo{journal}{Nature Nanotechnology}}
  \textbf{\bibinfo{volume}{8}}, \bibinfo{pages}{807} (\bibinfo{year}{2013}).

\bibitem{jones2015Optical}
\bibinfo{author}{Jones, P.~H.}, \bibinfo{author}{Marag{\`o}, O.~M.} \&
  \bibinfo{author}{Volpe, G.}
\newblock \emph{\bibinfo{title}{Optical Tweezers: Principles and Applications}}
  (\bibinfo{publisher}{Cambridge University Press}, \bibinfo{year}{2015}).

\bibitem{favre2019optical}
\bibinfo{author}{Favre-Bulle, I.~A.}, \bibinfo{author}{Stilgoe, A.~B.},
  \bibinfo{author}{Scott, E.~K.} \& \bibinfo{author}{Rubinsztein-Dunlop, H.}
\newblock \bibinfo{title}{Optical trapping in vivo: theory, practice, and
  applications}.
\newblock \emph{\bibinfo{journal}{Nanophotonics}} \textbf{\bibinfo{volume}{8}},
  \bibinfo{pages}{1023--1040} (\bibinfo{year}{2019}).

\bibitem{bustamante2021optical}
\bibinfo{author}{Bustamante, C.~J.}, \bibinfo{author}{Chemla, Y.~R.},
  \bibinfo{author}{Liu, S.} \& \bibinfo{author}{Wang, M.~D.}
\newblock \bibinfo{title}{Optical tweezers in single-molecule biophysics}.
\newblock \emph{\bibinfo{journal}{Nature Reviews Methods Primers}}
  \textbf{\bibinfo{volume}{1}}, \bibinfo{pages}{25} (\bibinfo{year}{2021}).

\bibitem{ashkin1986Observation}
\bibinfo{author}{Ashkin, A.}, \bibinfo{author}{Dziedzic, J.},
  \bibinfo{author}{Bjorkholm, J.} \& \bibinfo{author}{Chu, S.}
\newblock \bibinfo{title}{Observation of a single-beam gradient force optical
  trap for dielectric particles}.
\newblock \emph{\bibinfo{journal}{Optics Letters}}
  \textbf{\bibinfo{volume}{11}}, \bibinfo{pages}{288--290}
  (\bibinfo{year}{1986}).

\bibitem{block1990bead}
\bibinfo{author}{Block, S.~M.}, \bibinfo{author}{Goldstein, L.~S.} \&
  \bibinfo{author}{Schnapp, B.~J.}
\newblock \bibinfo{title}{Bead movement by single kinesin molecules studied
  with optical tweezers}.
\newblock \emph{\bibinfo{journal}{Nature}} \textbf{\bibinfo{volume}{348}},
  \bibinfo{pages}{348--352} (\bibinfo{year}{1990}).

\bibitem{heller2014optical}
\bibinfo{author}{Heller, I.}, \bibinfo{author}{Hoekstra, T.~P.},
  \bibinfo{author}{King, G.~A.}, \bibinfo{author}{Peterman, E.~J.} \&
  \bibinfo{author}{Wuite, G.~J.}
\newblock \bibinfo{title}{Optical tweezers analysis of dna--protein complexes}.
\newblock \emph{\bibinfo{journal}{Chemical Reviews}}
  \textbf{\bibinfo{volume}{114}}, \bibinfo{pages}{3087--3119}
  (\bibinfo{year}{2014}).

\bibitem{grier2003revolution}
\bibinfo{author}{Grier, D.~G.}
\newblock \bibinfo{title}{A revolution in optical manipulation}.
\newblock \emph{\bibinfo{journal}{Nature}} \textbf{\bibinfo{volume}{424}},
  \bibinfo{pages}{810--816} (\bibinfo{year}{2003}).

\bibitem{butaite2019indirect}
\bibinfo{author}{B{\=u}tait{\.e}, U.~G.} \emph{et~al.}
\newblock \bibinfo{title}{Indirect optical trapping using light driven
  micro-rotors for reconfigurable hydrodynamic manipulation}.
\newblock \emph{\bibinfo{journal}{Nature Communications}}
  \textbf{\bibinfo{volume}{10}}, \bibinfo{pages}{1215} (\bibinfo{year}{2019}).

\bibitem{berut2012experimental}
\bibinfo{author}{B{\'e}rut, A.} \emph{et~al.}
\newblock \bibinfo{title}{Experimental verification of landauer’s principle
  linking information and thermodynamics}.
\newblock \emph{\bibinfo{journal}{Nature}} \textbf{\bibinfo{volume}{483}},
  \bibinfo{pages}{187--189} (\bibinfo{year}{2012}).

\bibitem{saha2021maximizing}
\bibinfo{author}{Saha, T.~K.}, \bibinfo{author}{Lucero, J.~N.},
  \bibinfo{author}{Ehrich, J.}, \bibinfo{author}{Sivak, D.~A.} \&
  \bibinfo{author}{Bechhoefer, J.}
\newblock \bibinfo{title}{Maximizing power and velocity of an information
  engine}.
\newblock \emph{\bibinfo{journal}{Proceedings of the National Academy of
  Sciences}} \textbf{\bibinfo{volume}{118}}, \bibinfo{pages}{e2023356118}
  (\bibinfo{year}{2021}).

\bibitem{delic2020cooling}
\bibinfo{author}{Deli{\'c}, U.} \emph{et~al.}
\newblock \bibinfo{title}{Cooling of a levitated nanoparticle to the motional
  quantum ground state}.
\newblock \emph{\bibinfo{journal}{Science}} \textbf{\bibinfo{volume}{367}},
  \bibinfo{pages}{892--895} (\bibinfo{year}{2020}).

\bibitem{tebbenjohanns2021quantum}
\bibinfo{author}{Tebbenjohanns, F.}, \bibinfo{author}{Mattana, M.~L.},
  \bibinfo{author}{Rossi, M.}, \bibinfo{author}{Frimmer, M.} \&
  \bibinfo{author}{Novotny, L.}
\newblock \bibinfo{title}{Quantum control of a nanoparticle optically levitated
  in cryogenic free space}.
\newblock \emph{\bibinfo{journal}{Nature}} \textbf{\bibinfo{volume}{595}},
  \bibinfo{pages}{378--382} (\bibinfo{year}{2021}).

\bibitem{simpson2014inhomogeneous}
\bibinfo{author}{Simpson, S.}
\newblock \bibinfo{title}{Inhomogeneous and anisotropic particles in optical
  traps: Physical behaviour and applications}.
\newblock \emph{\bibinfo{journal}{Journal of Quantitative Spectroscopy and
  Radiative Transfer}} \textbf{\bibinfo{volume}{146}}, \bibinfo{pages}{81--99}
  (\bibinfo{year}{2014}).

\bibitem{blazquez2019optical}
\bibinfo{author}{Bl{\'a}zquez-Castro, A.}
\newblock \bibinfo{title}{Optical tweezers: Phototoxicity and thermal stress in
  cells and biomolecules}.
\newblock \emph{\bibinfo{journal}{Micromachines}}
  \textbf{\bibinfo{volume}{10}}, \bibinfo{pages}{507} (\bibinfo{year}{2019}).

\bibitem{peterman2003laser}
\bibinfo{author}{Peterman, E.~J.}, \bibinfo{author}{Gittes, F.} \&
  \bibinfo{author}{Schmidt, C.~F.}
\newblock \bibinfo{title}{Laser-induced heating in optical traps}.
\newblock \emph{\bibinfo{journal}{Biophysical Journal}}
  \textbf{\bibinfo{volume}{84}}, \bibinfo{pages}{1308--1316}
  (\bibinfo{year}{2003}).

\bibitem{friese1998optical}
\bibinfo{author}{Friese, M.}, \bibinfo{author}{Nieminen, T.},
  \bibinfo{author}{Heckenberg, N.} \& \bibinfo{author}{Rubinsztein-Dunlop, H.}
\newblock \bibinfo{title}{Optical alignment and spinning of laser-trapped
  microscopic particles}.
\newblock \emph{\bibinfo{journal}{Nature}} \textbf{\bibinfo{volume}{394}},
  \bibinfo{pages}{348} (\bibinfo{year}{1998}).

\bibitem{arlt2000generation}
\bibinfo{author}{Arlt, J.} \& \bibinfo{author}{Padgett, M.}
\newblock \bibinfo{title}{Generation of a beam with a dark focus surrounded by
  regions of higher intensity: the optical bottle beam}.
\newblock \emph{\bibinfo{journal}{Optics Letters}}
  \textbf{\bibinfo{volume}{25}}, \bibinfo{pages}{191--193}
  (\bibinfo{year}{2000}).

\bibitem{friese2001optically}
\bibinfo{author}{Friese, M.}, \bibinfo{author}{Rubinsztein-Dunlop, H.},
  \bibinfo{author}{Gold, J.}, \bibinfo{author}{Hagberg, P.} \&
  \bibinfo{author}{Hanstorp, D.}
\newblock \bibinfo{title}{Optically driven micromachine elements}.
\newblock \emph{\bibinfo{journal}{Applied Physics Letters}}
  \textbf{\bibinfo{volume}{78}}, \bibinfo{pages}{547--549}
  (\bibinfo{year}{2001}).

\bibitem{ladavac2004sorting}
\bibinfo{author}{Ladavac, K.}, \bibinfo{author}{Kasza, K.} \&
  \bibinfo{author}{Grier, D.~G.}
\newblock \bibinfo{title}{Sorting mesoscopic objects with periodic potential
  landscapes: Optical fractionation}.
\newblock \emph{\bibinfo{journal}{Physical Review E}}
  \textbf{\bibinfo{volume}{70}}, \bibinfo{pages}{010901}
  (\bibinfo{year}{2004}).

\bibitem{roichman2008optical}
\bibinfo{author}{Roichman, Y.}, \bibinfo{author}{Sun, B.},
  \bibinfo{author}{Roichman, Y.}, \bibinfo{author}{Amato-Grill, J.} \&
  \bibinfo{author}{Grier, D.~G.}
\newblock \bibinfo{title}{Optical forces arising from phase gradients}.
\newblock \emph{\bibinfo{journal}{Physical Review Letters}}
  \textbf{\bibinfo{volume}{100}}, \bibinfo{pages}{013602}
  (\bibinfo{year}{2008}).

\bibitem{dholakia2011shaping}
\bibinfo{author}{Dholakia, K.} \& \bibinfo{author}{{\v{C}}i{\v{z}}m{\'a}r, T.}
\newblock \bibinfo{title}{Shaping the future of manipulation}.
\newblock \emph{\bibinfo{journal}{Nature Photonics}}
  \textbf{\bibinfo{volume}{5}}, \bibinfo{pages}{335--342}
  (\bibinfo{year}{2011}).

\bibitem{padgett2011tweezers}
\bibinfo{author}{Padgett, M.} \& \bibinfo{author}{Bowman, R.}
\newblock \bibinfo{title}{Tweezers with a twist}.
\newblock \emph{\bibinfo{journal}{Nature Photonics}}
  \textbf{\bibinfo{volume}{5}}, \bibinfo{pages}{343} (\bibinfo{year}{2011}).

\bibitem{brzobohaty2013experimental}
\bibinfo{author}{Brzobohat{\`y}, O.} \emph{et~al.}
\newblock \bibinfo{title}{Experimental demonstration of optical transport,
  sorting and self-arrangement using a ‘tractor beam’}.
\newblock \emph{\bibinfo{journal}{Nature Photonics}}
  \textbf{\bibinfo{volume}{7}}, \bibinfo{pages}{123--127}
  (\bibinfo{year}{2013}).

\bibitem{woerdemann2013advanced}
\bibinfo{author}{Woerdemann, M.}, \bibinfo{author}{Alpmann, C.},
  \bibinfo{author}{Esseling, M.} \& \bibinfo{author}{Denz, C.}
\newblock \bibinfo{title}{Advanced optical trapping by complex beam shaping}.
\newblock \emph{\bibinfo{journal}{Laser \& Photonics Reviews}}
  \textbf{\bibinfo{volume}{7}}, \bibinfo{pages}{839--854}
  (\bibinfo{year}{2013}).

\bibitem{taylor2015enhanced}
\bibinfo{author}{Taylor, M.~A.}, \bibinfo{author}{Waleed, M.},
  \bibinfo{author}{Stilgoe, A.~B.}, \bibinfo{author}{Rubinsztein-Dunlop, H.} \&
  \bibinfo{author}{Bowen, W.~P.}
\newblock \bibinfo{title}{Enhanced optical trapping via structured scattering}.
\newblock \emph{\bibinfo{journal}{Nature Photonics}}
  \textbf{\bibinfo{volume}{9}}, \bibinfo{pages}{669} (\bibinfo{year}{2015}).

\bibitem{bezryadina2016optical}
\bibinfo{author}{Bezryadina, A.~S.}, \bibinfo{author}{Preece, D.~C.},
  \bibinfo{author}{Chen, J.~C.} \& \bibinfo{author}{Chen, Z.}
\newblock \bibinfo{title}{Optical disassembly of cellular clusters by tunable
  ‘tug-of-war’ tweezers}.
\newblock \emph{\bibinfo{journal}{Light: Science \& Applications}}
  \textbf{\bibinfo{volume}{5}}, \bibinfo{pages}{e16158--e16158}
  (\bibinfo{year}{2016}).

\bibitem{stilgoe2022controlled}
\bibinfo{author}{Stilgoe, A.~B.}, \bibinfo{author}{Nieminen, T.~A.} \&
  \bibinfo{author}{Rubinsztein-Dunlop, H.}
\newblock \bibinfo{title}{Controlled transfer of transverse orbital angular
  momentum to optically trapped birefringent microparticles}.
\newblock \emph{\bibinfo{journal}{Nature Photonics}}
  \textbf{\bibinfo{volume}{16}}, \bibinfo{pages}{346--351}
  (\bibinfo{year}{2022}).

\bibitem{hu2022structured}
\bibinfo{author}{Hu, Y.} \emph{et~al.}
\newblock \bibinfo{title}{Structured transverse orbital angular momentum probed
  by a levitated optomechanical sensor}.
\newblock \emph{\bibinfo{journal}{arXiv preprint arXiv:2209.09759}}
  (\bibinfo{year}{2022}).

\bibitem{friese1996determination}
\bibinfo{author}{Friese, M.}, \bibinfo{author}{Rubinsztein-Dunlop, H.},
  \bibinfo{author}{Heckenberg, N.} \& \bibinfo{author}{Dearden, E.}
\newblock \bibinfo{title}{Determination of the force constant of a single-beam
  gradient trap by measurement of backscattered light}.
\newblock \emph{\bibinfo{journal}{Applied Optics}}
  \textbf{\bibinfo{volume}{35}}, \bibinfo{pages}{7112--7116}
  (\bibinfo{year}{1996}).

\bibitem{simpson1998optical}
\bibinfo{author}{Simpson, N.}, \bibinfo{author}{McGloin, D.},
  \bibinfo{author}{Dholakia, K.}, \bibinfo{author}{Allen, L.} \&
  \bibinfo{author}{Padgett, M.}
\newblock \bibinfo{title}{Optical tweezers with increased axial trapping
  efficiency}.
\newblock \emph{\bibinfo{journal}{Journal of Modern Optics}}
  \textbf{\bibinfo{volume}{45}}, \bibinfo{pages}{1943--1949}
  (\bibinfo{year}{1998}).

\bibitem{bowman2010particle}
\bibinfo{author}{Bowman, R.}, \bibinfo{author}{Gibson, G.} \&
  \bibinfo{author}{Padgett, M.}
\newblock \bibinfo{title}{Particle tracking stereomicroscopy in optical
  tweezers: control of trap shape}.
\newblock \emph{\bibinfo{journal}{Optics Express}}
  \textbf{\bibinfo{volume}{18}}, \bibinfo{pages}{11785--11790}
  (\bibinfo{year}{2010}).

\bibitem{phillips2011optimizing}
\bibinfo{author}{Phillips, D.~B.} \emph{et~al.}
\newblock \bibinfo{title}{Optimizing the optical trapping stiffness of
  holographically trapped microrods using high-speed video tracking}.
\newblock \emph{\bibinfo{journal}{Journal of Optics}}
  \textbf{\bibinfo{volume}{13}}, \bibinfo{pages}{044023}
  (\bibinfo{year}{2011}).

\bibitem{mcalinden2014accurate}
\bibinfo{author}{McAlinden, N.}, \bibinfo{author}{Glass, D.~G.},
  \bibinfo{author}{Millington, O.~R.} \& \bibinfo{author}{Wright, A.~J.}
\newblock \bibinfo{title}{Accurate position tracking of optically trapped live
  cells}.
\newblock \emph{\bibinfo{journal}{Biomedical Optics Express}}
  \textbf{\bibinfo{volume}{5}}, \bibinfo{pages}{1026} (\bibinfo{year}{2014}).

\bibitem{singh2017particle}
\bibinfo{author}{Singh, B.~K.}, \bibinfo{author}{Nagar, H.},
  \bibinfo{author}{Roichman, Y.} \& \bibinfo{author}{Arie, A.}
\newblock \bibinfo{title}{Particle manipulation beyond the diffraction limit
  using structured super-oscillating light beams}.
\newblock \emph{\bibinfo{journal}{Light: Science \& Applications}}
  \textbf{\bibinfo{volume}{6}}, \bibinfo{pages}{e17050--e17050}
  (\bibinfo{year}{2017}).

\bibitem{mazilu2011optical}
\bibinfo{author}{Mazilu, M.}, \bibinfo{author}{Baumgartl, J.},
  \bibinfo{author}{Kosmeier, S.} \& \bibinfo{author}{Dholakia, K.}
\newblock \bibinfo{title}{Optical eigenmodes; exploiting the quadratic nature
  of the energy flux and of scattering interactions}.
\newblock \emph{\bibinfo{journal}{Optics Express}}
  \textbf{\bibinfo{volume}{19}}, \bibinfo{pages}{933--945}
  (\bibinfo{year}{2011}).

\bibitem{taylor2017optimizing}
\bibinfo{author}{Taylor, M.~A.}
\newblock \bibinfo{title}{Optimizing phase to enhance optical trap stiffness}.
\newblock \emph{\bibinfo{journal}{Scientific Reports}}
  \textbf{\bibinfo{volume}{7}}, \bibinfo{pages}{555} (\bibinfo{year}{2017}).

\bibitem{agate2004femtosecond}
\bibinfo{author}{Agate, B.}, \bibinfo{author}{Brown, C.},
  \bibinfo{author}{Sibbett, W.} \& \bibinfo{author}{Dholakia, K.}
\newblock \bibinfo{title}{Femtosecond optical tweezers for in-situ control of
  two-photon fluorescence}.
\newblock \emph{\bibinfo{journal}{Optics Express}}
  \textbf{\bibinfo{volume}{12}}, \bibinfo{pages}{3011--3017}
  (\bibinfo{year}{2004}).

\bibitem{ambichl2017focusing}
\bibinfo{author}{Ambichl, P.} \emph{et~al.}
\newblock \bibinfo{title}{Focusing inside disordered media with the generalized
  wigner-smith operator}.
\newblock \emph{\bibinfo{journal}{Physical Review Letters}}
  \textbf{\bibinfo{volume}{119}}, \bibinfo{pages}{033903}
  (\bibinfo{year}{2017}).

\bibitem{horodynski2020optimal}
\bibinfo{author}{Horodynski, M.} \emph{et~al.}
\newblock \bibinfo{title}{Optimal wave fields for micromanipulation in complex
  scattering environments}.
\newblock \emph{\bibinfo{journal}{Nature Photonics}}
  \textbf{\bibinfo{volume}{14}}, \bibinfo{pages}{149--153}
  (\bibinfo{year}{2020}).

\bibitem{volpe2023roadmap}
\bibinfo{author}{Volpe, G.} \emph{et~al.}
\newblock \bibinfo{title}{Roadmap for optical tweezers 2023 (chapter 2)}.
\newblock \emph{\bibinfo{journal}{Journal of Physics: Photonics}}
  (\bibinfo{year}{2023}).

\bibitem{simpson2009thermal}
\bibinfo{author}{Simpson, S.~H.} \& \bibinfo{author}{Hanna, S.}
\newblock \bibinfo{title}{Thermal motion of a holographically trapped spm-like
  probe}.
\newblock \emph{\bibinfo{journal}{Nanotechnology}}
  \textbf{\bibinfo{volume}{20}}, \bibinfo{pages}{395710}
  (\bibinfo{year}{2009}).

\bibitem{phillips2012optically}
\bibinfo{author}{Phillips, D.} \emph{et~al.}
\newblock \bibinfo{title}{An optically actuated surface scanning probe}.
\newblock \emph{\bibinfo{journal}{Optics Express}}
  \textbf{\bibinfo{volume}{20}}, \bibinfo{pages}{29679} (\bibinfo{year}{2012}).

\bibitem{phillips2012force}
\bibinfo{author}{Phillips, D.~B.} \emph{et~al.}
\newblock \bibinfo{title}{Force sensing with a shaped dielectric micro-tool}.
\newblock \emph{\bibinfo{journal}{EPL (Europhysics Letters)}}
  \textbf{\bibinfo{volume}{99}}, \bibinfo{pages}{58004} (\bibinfo{year}{2012}).

\bibitem{waterman1971symmetry}
\bibinfo{author}{Waterman, P.~C.}
\newblock \bibinfo{title}{Symmetry, unitarity, and geometry in electromagnetic
  scattering}.
\newblock \emph{\bibinfo{journal}{Physical Review D}}
  \textbf{\bibinfo{volume}{3}}, \bibinfo{pages}{825} (\bibinfo{year}{1971}).

\bibitem{nieminen2007optical}
\bibinfo{author}{Nieminen, T.~A.} \emph{et~al.}
\newblock \bibinfo{title}{Optical tweezers computational toolbox}.
\newblock \emph{\bibinfo{journal}{Journal of Optics A: Pure and Applied
  Optics}} \textbf{\bibinfo{volume}{9}}, \bibinfo{pages}{S196}
  (\bibinfo{year}{2007}).

\bibitem{barton1989theoretical}
\bibinfo{author}{Barton, J.}, \bibinfo{author}{Alexander, D.} \&
  \bibinfo{author}{Schaub, S.}
\newblock \bibinfo{title}{Theoretical determination of net radiation force and
  torque for a spherical particle illuminated by a focused laser beam}.
\newblock \emph{\bibinfo{journal}{Journal of Applied Physics}}
  \textbf{\bibinfo{volume}{66}}, \bibinfo{pages}{4594--4602}
  (\bibinfo{year}{1989}).

\bibitem{butaite2020enhanced}
\bibinfo{author}{B{\=u}tait{\.e}, U.~G.}
\newblock \emph{\bibinfo{title}{Enhanced optical tweezing: from hydrodynamic
  micro-manipulation to optimised optical trapping}}.
\newblock Ph.D. thesis, \bibinfo{school}{University of Glasgow}
  (\bibinfo{year}{2020}).

\bibitem{wigner1955lower}
\bibinfo{author}{Wigner, E.~P.}
\newblock \bibinfo{title}{Lower limit for the energy derivative of the
  scattering phase shift}.
\newblock \emph{\bibinfo{journal}{Physical Review}}
  \textbf{\bibinfo{volume}{98}}, \bibinfo{pages}{145} (\bibinfo{year}{1955}).

\bibitem{smith1960lifetime}
\bibinfo{author}{Smith, F.~T.}
\newblock \bibinfo{title}{Lifetime matrix in collision theory}.
\newblock \emph{\bibinfo{journal}{Physical Review}}
  \textbf{\bibinfo{volume}{118}}, \bibinfo{pages}{349} (\bibinfo{year}{1960}).

\bibitem{phillips2014shape}
\bibinfo{author}{Phillips, D.} \emph{et~al.}
\newblock \bibinfo{title}{Shape-induced force fields in optical trapping}.
\newblock \emph{\bibinfo{journal}{Nature Photonics}}
  \textbf{\bibinfo{volume}{8}}, \bibinfo{pages}{400} (\bibinfo{year}{2014}).

\bibitem{hay2014four}
\bibinfo{author}{Hay, R.}, \bibinfo{author}{Gibson, G.}, \bibinfo{author}{Lee,
  M.}, \bibinfo{author}{Padgett, M.} \& \bibinfo{author}{Phillips, D.}
\newblock \bibinfo{title}{Four-directional stereo-microscopy for 3d particle
  tracking with real-time error evaluation}.
\newblock \emph{\bibinfo{journal}{Optics Express}}
  \textbf{\bibinfo{volume}{22}}, \bibinfo{pages}{18662--18667}
  (\bibinfo{year}{2014}).

\bibitem{moser2019model}
\bibinfo{author}{Moser, S.}, \bibinfo{author}{Ritsch-Marte, M.} \&
  \bibinfo{author}{Thalhammer, G.}
\newblock \bibinfo{title}{Model-based compensation of pixel crosstalk in liquid
  crystal spatial light modulators}.
\newblock \emph{\bibinfo{journal}{Optics Express}}
  \textbf{\bibinfo{volume}{27}}, \bibinfo{pages}{25046--25063}
  (\bibinfo{year}{2019}).

\bibitem{kupianskyi2023high}
\bibinfo{author}{Kupianskyi, H.}, \bibinfo{author}{Horsley, S.~A.} \&
  \bibinfo{author}{Phillips, D.~B.}
\newblock \bibinfo{title}{High-dimensional spatial mode sorting and optical
  circuit design using multi-plane light conversion}.
\newblock \emph{\bibinfo{journal}{APL Photonics}} \textbf{\bibinfo{volume}{8}},
  \bibinfo{pages}{026101} (\bibinfo{year}{2023}).

\bibitem{gu1996effect}
\bibinfo{author}{Gu, M.}
\newblock \bibinfo{title}{Effect of apodization on axial resolution with a
  high-aperture objective}.
\newblock In \emph{\bibinfo{booktitle}{Three-Dimensional Microscopy: Image
  Acquisition and Processing III}}, vol. \bibinfo{volume}{2655},
  \bibinfo{pages}{53--61} (\bibinfo{organization}{SPIE}, \bibinfo{year}{1996}).

\bibitem{sheppard1988aberrations}
\bibinfo{author}{Sheppard, C.~J.}
\newblock \bibinfo{title}{Aberrations in high aperture conventional and
  confocal imaging systems}.
\newblock \emph{\bibinfo{journal}{Applied Optics}}
  \textbf{\bibinfo{volume}{27}}, \bibinfo{pages}{4782--4786}
  (\bibinfo{year}{1988}).

\bibitem{he2021vectorial}
\bibinfo{author}{He, C.}, \bibinfo{author}{Antonello, J.} \&
  \bibinfo{author}{Booth, M.~J.}
\newblock \bibinfo{title}{Vectorial adaptive optics}.
\newblock \emph{\bibinfo{journal}{arXiv preprint arXiv:2110.02606}}
  (\bibinfo{year}{2021}).

\bibitem{iwaniuk2011correcting}
\bibinfo{author}{Iwaniuk, D.}, \bibinfo{author}{Rastogi, P.} \&
  \bibinfo{author}{Hack, E.}
\newblock \bibinfo{title}{Correcting spherical aberrations induced by an
  unknown medium through determination of its refractive index and thickness}.
\newblock \emph{\bibinfo{journal}{Optics Express}}
  \textbf{\bibinfo{volume}{19}}, \bibinfo{pages}{19407--19414}
  (\bibinfo{year}{2011}).

\bibitem{vcivzmar2010situ}
\bibinfo{author}{{\v{C}}i{\v{z}}m{\'a}r, T.}, \bibinfo{author}{Mazilu, M.} \&
  \bibinfo{author}{Dholakia, K.}
\newblock \bibinfo{title}{In situ wavefront correction and its application to
  micromanipulation}.
\newblock \emph{\bibinfo{journal}{Nature Photonics}}
  \textbf{\bibinfo{volume}{4}}, \bibinfo{pages}{388--394}
  (\bibinfo{year}{2010}).

\bibitem{vellekoop2007focusing}
\bibinfo{author}{Vellekoop, I.~M.} \& \bibinfo{author}{Mosk, A.}
\newblock \bibinfo{title}{Focusing coherent light through opaque strongly
  scattering media}.
\newblock \emph{\bibinfo{journal}{Optics Letters}}
  \textbf{\bibinfo{volume}{32}}, \bibinfo{pages}{2309--2311}
  (\bibinfo{year}{2007}).

\bibitem{gigan2022roadmap}
\bibinfo{author}{Gigan, S.} \emph{et~al.}
\newblock \bibinfo{title}{Roadmap on wavefront shaping and deep imaging in
  complex media}.
\newblock \emph{\bibinfo{journal}{Journal of Physics: Photonics}}
  \textbf{\bibinfo{volume}{4}}, \bibinfo{pages}{042501} (\bibinfo{year}{2022}).

\bibitem{cao2022shaping}
\bibinfo{author}{Cao, H.}, \bibinfo{author}{Mosk, A.~P.} \&
  \bibinfo{author}{Rotter, S.}
\newblock \bibinfo{title}{Shaping the propagation of light in complex media}.
\newblock \emph{\bibinfo{journal}{Nature Physics}}
  \textbf{\bibinfo{volume}{18}}, \bibinfo{pages}{994--1007}
  (\bibinfo{year}{2022}).

\bibitem{mastiani2021noise}
\bibinfo{author}{Mastiani, B.} \& \bibinfo{author}{Vellekoop, I.~M.}
\newblock \bibinfo{title}{Noise-tolerant wavefront shaping in a hadamard
  basis}.
\newblock \emph{\bibinfo{journal}{Optics Express}}
  \textbf{\bibinfo{volume}{29}}, \bibinfo{pages}{17534--17541}
  (\bibinfo{year}{2021}).

\bibitem{mididoddi2023threading}
\bibinfo{author}{Mididoddi, C.~K.}, \bibinfo{author}{Sharp, C.},
  \bibinfo{author}{del Hougne, P.}, \bibinfo{author}{Horsley, S.~A.} \&
  \bibinfo{author}{Phillips, D.~B.}
\newblock \bibinfo{title}{Threading light through dynamic complex media}.
\newblock \emph{\bibinfo{journal}{arXiv preprint arXiv:2301.04461}}
  (\bibinfo{year}{2023}).

\bibitem{ashkin1970acceleration}
\bibinfo{author}{Ashkin, A.}
\newblock \bibinfo{title}{Acceleration and trapping of particles by radiation
  pressure}.
\newblock \emph{\bibinfo{journal}{Physical Review Letters}}
  \textbf{\bibinfo{volume}{24}}, \bibinfo{pages}{156} (\bibinfo{year}{1970}).

\bibitem{thalhammer2011optical}
\bibinfo{author}{Thalhammer, G.}, \bibinfo{author}{Steiger, R.},
  \bibinfo{author}{Bernet, S.} \& \bibinfo{author}{Ritsch-Marte, M.}
\newblock \bibinfo{title}{Optical macro-tweezers: trapping of highly motile
  micro-organisms}.
\newblock \emph{\bibinfo{journal}{Journal of Optics}}
  \textbf{\bibinfo{volume}{13}}, \bibinfo{pages}{044024}
  (\bibinfo{year}{2011}).

\bibitem{xiao2021efficient}
\bibinfo{author}{Xiao, Y.} \emph{et~al.}
\newblock \bibinfo{title}{Efficient generation of optical bottle beams}.
\newblock \emph{\bibinfo{journal}{Nanophotonics}}
  \textbf{\bibinfo{volume}{10}}, \bibinfo{pages}{2893--2901}
  (\bibinfo{year}{2021}).

\bibitem{melo2020optical}
\bibinfo{author}{Melo, B.} \emph{et~al.}
\newblock \bibinfo{title}{Optical trapping in a dark focus}.
\newblock \emph{\bibinfo{journal}{Physical Review Applied}}
  \textbf{\bibinfo{volume}{14}}, \bibinfo{pages}{034069}
  (\bibinfo{year}{2020}).

\bibitem{lee2017computational}
\bibinfo{author}{Lee, Y.~E.}, \bibinfo{author}{Miller, O.~D.},
  \bibinfo{author}{Reid, M.~H.}, \bibinfo{author}{Johnson, S.~G.} \&
  \bibinfo{author}{Fang, N.~X.}
\newblock \bibinfo{title}{Computational inverse design of non-intuitive
  illumination patterns to maximize optical force or torque}.
\newblock \emph{\bibinfo{journal}{Optics Express}}
  \textbf{\bibinfo{volume}{25}}, \bibinfo{pages}{6757--6766}
  (\bibinfo{year}{2017}).

\bibitem{kim2017tomographic}
\bibinfo{author}{Kim, K.} \& \bibinfo{author}{Park, Y.}
\newblock \bibinfo{title}{Tomographic active optical trapping of arbitrarily
  shaped objects by exploiting 3d refractive index maps}.
\newblock \emph{\bibinfo{journal}{Nature Communications}}
  \textbf{\bibinfo{volume}{8}}, \bibinfo{pages}{1--8} (\bibinfo{year}{2017}).

\bibitem{landenberger2021nonblind}
\bibinfo{author}{Landenberger, B.}, \bibinfo{author}{{Yatish}} \&
  \bibinfo{author}{Rohrbach, A.}
\newblock \bibinfo{title}{Towards non-blind optical tweezing by finding 3d
  refractive index changes through off-focus interferometric tracking}.
\newblock \emph{\bibinfo{journal}{Nature Communications}}
  \textbf{\bibinfo{volume}{12}}, \bibinfo{pages}{6922} (\bibinfo{year}{2021}).

\bibitem{hu2008antireflection}
\bibinfo{author}{Hu, Y.}, \bibinfo{author}{Nieminen, T.~A.},
  \bibinfo{author}{Heckenberg, N.~R.} \& \bibinfo{author}{Rubinsztein-Dunlop,
  H.}
\newblock \bibinfo{title}{Antireflection coating for improved optical
  trapping}.
\newblock \emph{\bibinfo{journal}{Journal of Applied Physics}}
  \textbf{\bibinfo{volume}{103}}, \bibinfo{pages}{093119}
  (\bibinfo{year}{2008}).

\bibitem{jannasch2012nanonewton}
\bibinfo{author}{Jannasch, A.}, \bibinfo{author}{Demir{\"o}rs, A.~F.},
  \bibinfo{author}{Van~Oostrum, P.~D.}, \bibinfo{author}{Van~Blaaderen, A.} \&
  \bibinfo{author}{Sch{\"a}ffer, E.}
\newblock \bibinfo{title}{Nanonewton optical force trap employing
  anti-reflection coated, high-refractive-index titania microspheres}.
\newblock \emph{\bibinfo{journal}{Nature Photonics}}
  \textbf{\bibinfo{volume}{6}}, \bibinfo{pages}{469--473}
  (\bibinfo{year}{2012}).

\bibitem{piotrowski2023simultaneous}
\bibinfo{author}{Piotrowski, J.} \emph{et~al.}
\newblock \bibinfo{title}{Simultaneous ground-state cooling of two mechanical
  modes of a levitated nanoparticle}.
\newblock \emph{\bibinfo{journal}{Nature Physics}} \bibinfo{pages}{1--5}
  (\bibinfo{year}{2023}).

\bibitem{millen2014nanoscale}
\bibinfo{author}{Millen, J.}, \bibinfo{author}{Deesuwan, T.},
  \bibinfo{author}{Barker, P.} \& \bibinfo{author}{Anders, J.}
\newblock \bibinfo{title}{Nanoscale temperature measurements using
  non-equilibrium brownian dynamics of a levitated nanosphere}.
\newblock \emph{\bibinfo{journal}{Nature Nanotechnology}}
  \textbf{\bibinfo{volume}{9}}, \bibinfo{pages}{425--429}
  (\bibinfo{year}{2014}).

\bibitem{melzer2021assembly}
\bibinfo{author}{Melzer, J.~E.} \& \bibinfo{author}{McLeod, E.}
\newblock \bibinfo{title}{Assembly of multicomponent structures from hundreds
  of micron-scale building blocks using optical tweezers}.
\newblock \emph{\bibinfo{journal}{Microsystems \& Nanoengineering}}
  \textbf{\bibinfo{volume}{7}}, \bibinfo{pages}{45} (\bibinfo{year}{2021}).

\bibitem{perez2018high}
\bibinfo{author}{P{\'e}rez~Garc{\'\i}a, L.},
  \bibinfo{author}{Donlucas~P{\'e}rez, J.}, \bibinfo{author}{Volpe, G.},
  \bibinfo{author}{V.~Arzola, A.} \& \bibinfo{author}{Volpe, G.}
\newblock \bibinfo{title}{High-performance reconstruction of microscopic force
  fields from brownian trajectories}.
\newblock \emph{\bibinfo{journal}{Nature Communications}}
  \textbf{\bibinfo{volume}{9}}, \bibinfo{pages}{5166} (\bibinfo{year}{2018}).

\bibitem{gibson2012compact}
\bibinfo{author}{Gibson, G.} \emph{et~al.}
\newblock \bibinfo{title}{A compact holographic optical tweezers instrument}.
\newblock \emph{\bibinfo{journal}{Review of Scientific Instruments}}
  \textbf{\bibinfo{volume}{83}}, \bibinfo{pages}{113107}
  (\bibinfo{year}{2012}).

\bibitem{bowman2014red}
\bibinfo{author}{Bowman, R.~W.} \emph{et~al.}
\newblock \bibinfo{title}{Red tweezers: Fast, customisable hologram generation
  for optical tweezers}.
\newblock \emph{\bibinfo{journal}{Computer Physics Communications}}
  \textbf{\bibinfo{volume}{185}}, \bibinfo{pages}{268--273}
  (\bibinfo{year}{2014}).

\bibitem{crameri2020misuse}
\bibinfo{author}{Crameri, F.}, \bibinfo{author}{Shephard, G.~E.} \&
  \bibinfo{author}{Heron, P.~J.}
\newblock \bibinfo{title}{The misuse of colour in science communication}.
\newblock \emph{\bibinfo{journal}{Nature Communications}}
  \textbf{\bibinfo{volume}{11}}, \bibinfo{pages}{1--10} (\bibinfo{year}{2020}).

\bibitem{crameri2021scientific}
\bibinfo{author}{Crameri, F.}
\newblock \bibinfo{title}{Scientific colour maps}.
\newblock \bibinfo{howpublished}{\url{https://doi.org/10.5281/zenodo.5501399}}
  (\bibinfo{year}{2021}).
\newblock \bibinfo{note}{Retrieved April 17, 2023}.

\bibitem{tim2021voxview}
\bibinfo{author}{Lukas}.
\newblock \bibinfo{title}{Voxview}.
\newblock
  \bibinfo{howpublished}{\url{https://www.mathworks.com/matlabcentral/fileexchange/78745-voxview}}
  (\bibinfo{year}{2021}).
\newblock \bibinfo{note}{MATLAB Central File Exchange. Retrieved April 17,
  2023}.

\end{thebibliography}

\end{document}